\documentclass[a4paper,10pt,twocolumn]{article}
\usepackage{newtxtext,newtxmath}
\usepackage{amsmath}
\usepackage{graphicx}
\usepackage[margin=17mm,a4paper]{geometry}
\usepackage[final]{hyperref}
\hypersetup{colorlinks=true,linkcolor=blue,citecolor=blue,filecolor=magenta,urlcolor=blue}

\begin{document}

\title{\Large\textbf{The accuracy of signal measurement with the water-Cherenkov detectors of the Pierre Auger Observatory}}
\author{\normalsize
M.~Ave,$^\text{a}$
P.~Bauleo,$^\text{b}$\footnotemark~
A.~Castellina,$^\text{c}$
A.~Chou,$^\text{d}$
J.L.~Harton,$^\text{b}$
R.~Knapik,$^\text{b}$
and
G.~Navarra,$^\text{c}$
\\\normalsize
for the Pierre Auger Collaboration
\\\footnotesize
$^\text{a}$\emph{Enrico Fermi Institute, University of Chicago, 5640 S. Ellis, Chicago, IL60637, USA}
\\[-1mm]\footnotesize
$^\text{b}$\emph{Department of Physics, Colorado State University, Fort Collins, CO 80523, USA}
\\[-1mm]\footnotesize
$^\text{c}$\emph{Dipartimento di Fisica Generale dell’Universit\`a, Istituto di Fisica dello Spazio Interplanetario (INAF) and}
\\[-2.5mm]\footnotesize
\emph{INFN, Torino, via P.~Giuria, 1, 10125 Torino, Italy} 
\\[-1mm]\footnotesize
$^\text{d}$\emph{Fermilab, MS367, POB 500, Batavia, IL 60510-0500, USA}
}
\date{}

\twocolumn[
\begin{@twocolumnfalse}
\maketitle
\vspace{-7mm}
~\\\hrule
\vspace{-1mm}
\begin{abstract}
The Auger Surface Detector consists of a large array of water-Cherenkov detector tanks each with a volume of $12,000\,\ell$, for the detection of high energy cosmic rays.
The accuracy in the measurement of the integrated signal amplitude of the detector unit has been studied using experimental air shower data.
It can be described as a Poisson-like term with a normalization constant that depends on the zenith angle of the primary cosmic ray.
This dependence reﬂects the increasing contribution to the signal of the muonic component of the shower, both due to the increasing muon/electromagnetic (e$^\pm$ and $\gamma$) ratio and muon track length with zenith angle.

~

\noindent
\emph{PACS}: 95.55.Vj; 96.50.sd

~

\noindent
\emph{Keywords}: Auger; Extensive air shower; Cherenkov detector; Cosmic ray

\end{abstract}
\hrule
\vspace{5mm}

Published in Nucl.~Instrum.~Meth.~A as \href{https://doi.org/10.1016/j.nima.2007.05.150}{DOI:10.1016/j.nima.2007.05.150}
\\
Report number: FERMILAB-PUB-07-681-E
\vspace{10mm}
\end{@twocolumnfalse}
]

\let\theorigfootnote\thefootnote
\renewcommand{\thefootnote}{*}\footnotetext{Corresponding author, \texttt{bauleo@lamar.colostate.edu}}
\let\thefootnote\theorigfootnote

\section{Introduction}

The Pierre Auger experiment calls for the construction of two large detector arrays, one in the southern hemisphere and one in the northern hemisphere, each covering an area of at least 3000\,km$^2$ to measure Extensive Air Showers (EAS) initiated by cosmic rays with energies above about $3{\times}10^{18}$\,eV.
The Southern Observatory is currently under construction and will consist of 1600 water-Cherenkov detectors located on a triangular array of 1.5\,km on a side to measure secondary particles reaching the ground.
In addition, 24 telescopes located on the array boundaries overlooking the southern array [1] will measure nitrogen ﬂuorescence light produced by the showers in the air on dark nights.

The arrival direction and energy of a cosmic ray are measured through its cascade of secondary particles.
The arrival direction can be deduced from the shower front arrival time at the different surface detectors (SDs).
The detector signals also provide information on the barycenter impact position at ground, or core location, and lateral spread of the shower or \emph{Lateral Distribution Function} (LDF).
The signal measured or interpolated at 1000\,m (in the shower plane) from the core, usually referred as $S(1000)$, can be related to the primary energy (see Ref.~[2] in which due to the geometric conﬁguration optimized for lower energies $S(600)$ was used).
The Auger Surface Detector unit consists of a completely enclosed 12,000-$\ell$ cylindrical tank of ultra pure water.
The footprint of the detector volume is 10\,m$^2$.
The water is observed by three 9-in photomultiplier tubes (PMT) which record the Cherenkov light produced when secondary particles of the cascade (mainly muons, electrons and gammas) traverse the water volume [3,4].

 This study is  focused  on  the  measurement  accuracy  of the integrated signal amplitudes, i.e.\ of the energy deposited in the individual SD stations.
 Such accuracy is a basic element in the reconstruction procedure, and has to be determined experimentally using real events measured with the detector. Shower ﬂuctuations are extremely difﬁcult to simulate, due to the large number of particles of the cascades (${>}10^{11}$) and the uncertainties in the numbers   of   muons,   electrons   and   gamma-rays, which depend on the hadron interactions and on the primary cosmic ray particle type.
 Moreover, the measurements in include the calibrations and the long term (months, years) gain monitoring procedures.

The electronics of each detector sample and digitize the signals produced by each PMT every 25\,ns.
Although the rich time structure of the signal carries detailed information of the shower, in this analysis the FADC trace is time-integrated and the signal is measured in units of VEM.
A VEM (Vertical Equivalent Muon) is deﬁned as the sum of the charge collected by the three PMTs when a single cosmic ray muon vertically traverses the detector moving downward along the detector axis [5].
All signals are measured in VEM units, regardless of the particle species crossing the detector.
Further details on the Auger experiment are available elsewhere [3,6].

In order to measure the SD signal accuracy two detectors have been deployed in the same position of the array, and this has been done at two nominal array positions, located 1.5\,km from each other.
In each one of these positions, the two tanks are located about 10\,m from each other.
As the footprint of an EAS is of the order of several square kilometers, these tanks are virtually measuring the same spot in the shower.
The positions have been named after the tanks located there.
Tanks named D{\'\i}a and Noche (DN) form one twin pair.
Located at a neighboring array position is the second pair, with the stations Moulin and Rouge (MR).

The detector signal accuracy depends on the measured signal ($S$) but it is also possible to expect differences due to other factors including core distance ($r$), shower zenith angle ($\theta$), or cosmic ray particle type.
We ﬁnd that the ratio of the number of electrons and gammas (EM component) to the number of muons in the shower affects the signal accuracy.
This ratio, and hence the signal accuracy, is affected by primary zenith angle since the electromagnetic particles are attenuated in the atmosphere more than muons.
The statistics used in the present analysis (January 2004 -- May 2006, about 2800 events) allow us to extract information on the signal accuracy both as a function of  the signal amplitude and the shower zenith angle.
The steep power-law energy spectrum of the cosmic ray implies that most of the recorded events in this analysis corresponds to energies of a few EeV.
These showers trigger the detector when the core is located roughly in the center of the basic equilateral  triangle  of  the  array,   therefore   most   of   the events have been recorded at a distance range of 600 to 900 m, not allowing us to extract detailed information on the signal accuracy dependence on core distance.

\section{Analysis method}

Having two detectors measuring basically the same spot on the shower allows us to measure the signal ﬂuctuation by analyzing the difference of their signals for a given shower.
Fig.~1  shows  the  signal  correlation  between  the detectors.
The relative signal ﬂuctuation has been deﬁned as
\begin{equation}
   \Delta \equiv \sqrt{2} \times \frac{(S_1-S_2)}{(S_1+S_2)}, 
   \label{Eq:equation1} 
\end{equation}
where $S_i$ corresponds to the signal  in the $i^\text{th}$ detector of the pair (in VEM units).
The $\sqrt{2}$ factor was added to take into account that $\Delta$ is defined as the relative signal ﬂuctuation between two tanks.
The relative signal accuracy of a single detector is then given by the width of the $\Delta$ distribution, $\delta \Delta=\sigma /S$ where $\sigma$ is the single detector accuracy.
In order to obtain this expression it has been assumed that $S_1 \sim S_2$ and that $\sigma_1 \sim \sigma_2$.
These assumptions are reasonable since quality cuts will ensure the detectors measure real EAS events, and the detectors are in principle identical, therefore their accuracies should be similar.
This has been observed as, within statistical errors, the two pairs give similar results.

\begin{figure}[t]
\centering
\includegraphics[width=0.95\columnwidth,angle=-90]{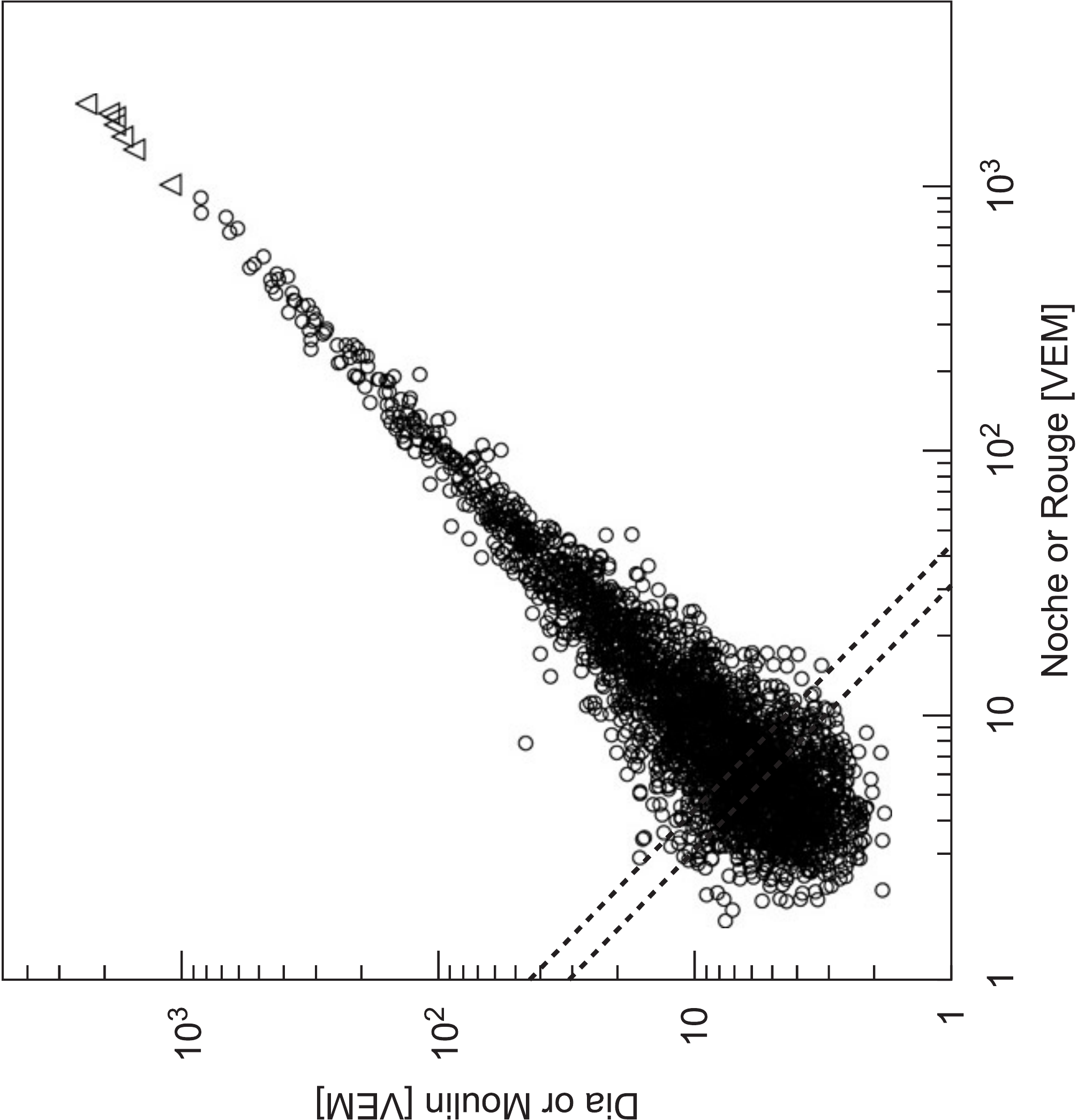}
\caption{Scatter plot of signals in the twin stations.
Triangles indicate the large amplitude signals where the electronics dynamic range was saturated.
The width of the correlation distribution, projected into different average signal bins is then used to obtain the detector signal accuracy.
The bin boundaries for one bin of constant average signal are shown by dotted lines.}
\end{figure}

The procedure to measure the detector accuracy starts by selecting events (event selection is described in Section 3) and binning them as function of the measured average signal.
Bins of measured average signal appear in Fig.~1 as slices perpendicular to the main correlation of the data.
The bin boundaries for one such bin are shown by dotted lines in Fig.~1.
The $\Delta$ distribution in this bin is just the histogram of the data centered on the average value, and the RMS of that distribution is the normalized single detector accuracy for that bin.
The bins at low average signal show larger width perpendicular to the main correlation, and hence the signal accuracy is worse there than at higher average signals where the data hug the diagonal more tightly.
The error on the normalized single detector accuracy for each bin was obtained from the variance of $(\sigma/S)^2$ [7]:
\begin{equation}
   V[{(\sigma/S)^2]} = \frac{1}{N} \left (m_4 - \frac {N-3}{N-1} {m_2^2}  \right),
   \label{Eq:equation2} 
\end{equation}
where $m_i$ is the $i^\text{th}$ central moment, i.e.\ $m_i = \frac {1}{N} \Sigma {(\Delta -\bar{\Delta})^i}$  and $N$ is  the  number  of  entries  of  the given bin.
This expression will be used to calculate the error bars in Fig.~2.
For a Gaussian distribution and large $N$ this yields the standard deviation of $\sigma/S : \sigma_\text{fit} / \sqrt{2N}$, where $\sigma_\text{fit}$  would be the Gaussian width obtained from a minimization procedure.

\begin{figure}[t]
\centering
\includegraphics[width=\columnwidth]{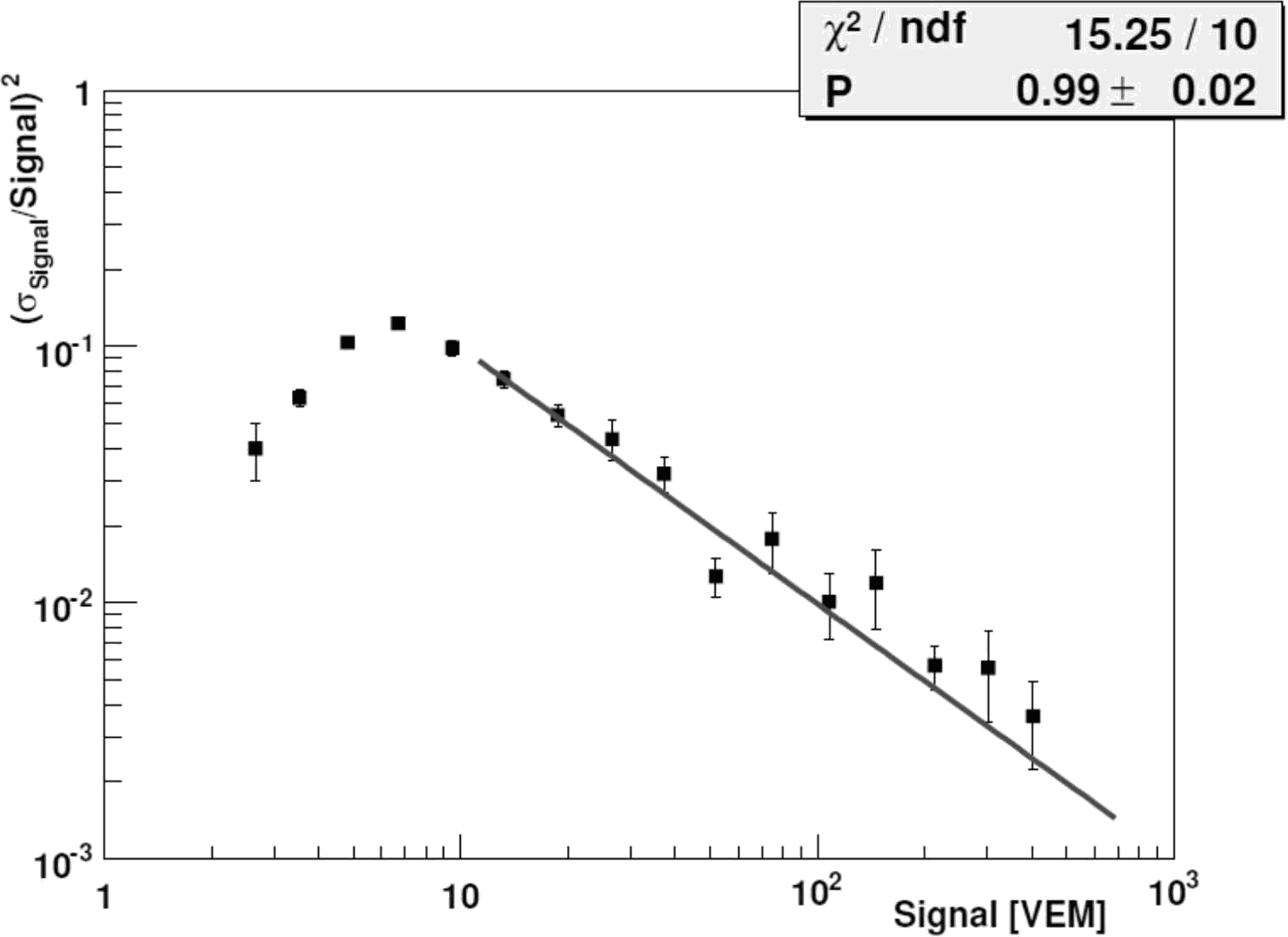}
\caption{Surface detector signal accuracy.
Values below 10\,VEM are distorted by triggering efﬁciency.
The ﬁt for a single parameter ($P$) is performed to the signal range unaffected by triggering.
The line represents a Poissonian-like ﬁt of the type $\sigma/\text{Signal}=P/\sqrt{\text{Signal}}$.}
\end{figure}

The relative difference of obtaining the error of $\sigma / S$ using Eq.~(2) or ﬁtting a Gaussian to the $\sigma / S$ distribution is bounded by 20\%, except for the large signal bins, where the total number of entries is small and the Gaussian approximation is not valid.
Due to the large signal bins (with relatively few entries) we decided to use Eq.~(2) in our error calculation.

\section{Signal accuracy measurements}

The selection requirements can be split into two categories, those that involve only tank status and those  that involve the event reconstruction:
\begin{itemize}
  \item \emph{Tank requirements}:
  
(1)	Three PMTs functioning properly in both tanks.

(2)	Non-saturated signal amplitude.

  \item \emph{Event reconstruction requirements}:
  
(3)	Successfully reconstructed event with three or more neighboring tank positions, after removal of random and isolated detectors (i.e.\ Auger Trigger level T4 [8]).

(4)	Reconstructed core position beyond 200\,m from both tanks
\end{itemize}

Requirement 1 guarantees a full operational unit and prevents our data sample to be biased due to the \emph {Azimuthal effect} (asymmetric charge collection when a PMT is not working [9]), while requirement 2 avoids electronics dynamic range problems.
Requirement 2 is applied over the summed signal of the three PMTs.
In requirement 3 the twin-tanks count as one position and selects well measured events.

Requirement 4 allows to neglect the 10\,m separation between the tanks and consider both of them at the same distance from the core.
For core positions less than 200\,m from the twin-tank location, the steepness of the LDF cannot be neglected (\emph {LDF Effect}) yielding a difference in the measured signal in the detectors.

Analysis of the signal accuracy shows that both tank pairs yield the same signal accuracy and their data have been merged.
A total of 2872 events were selected from the period January 1st 2004 to May 31st 2006.

\subsection {Data results}

Fig.~2 shows ${\sigma / S}^2$ versus $S$, which is the basic information from which we extract our results.
The signal accuracy gets distorted below about 10 VEM due to triggering effects (see Section 3.4).
To avoid biasing our results due to this effect, we ﬁt only the large signal range as shown in Fig.~2.
The signal bin size has been chosen as $\log_{10} (S_{max}/S_{min})=0.3$.
We require to each signal bin to have at least 10 entries.

We found that the measured signal ﬂuctuation (average value weighted by the zenith angle distribution ranging from 0$^{\circ}$ to 90$^{\circ}$  can be represented by:
\begin{equation}
   \sigma [\text{VEM}] = P \times \sqrt{S}.
   \label{Eq:equation3} 
\end{equation}
 A minimization  to the data yields  $P = (0.99 \pm 0.02)$   with a $\chi ^2$ value of 15/10 d.o.f.\ showing that the chosen simple and natural ``Poisson-like'' dependence represents an acceptable expression in this analysis.
 Error bars were obtained  using Eq.~(2).
 This expression is valid for non-saturated signals, beyond 200\,m from the shower core, which imposes a practical limit on the signal of about 500\,VEM.
 As mentioned before, most of the showers used in this analysis had energies around a few EeV.
 Small corrections to the previous expression could be expected for a different energy range

\subsection{Zenith angle dependence}

Within the limitations of the current statistics we have measured the zenith angle dependence of the signal accuracy.
The zenith angle dependence of the signal accuracy has been studied by repeating the previous analysis in four bins (from 0$^{\circ}$ to 68$^{\circ}$) of zenith angle.
A linear parametrization on $\sigma$ s with the secant of shower zenith angle was done and the ﬁtted values are (see Fig.~3):
\begin{equation}
   \sigma  = [(0.32 \pm 0.09) + (0.42 \pm 0.07) \times \sec(\theta)] \sqrt{s}.
   \label{Eq:equation4} 
\end{equation}
This zenith angle dependence is due mainly to the muonic component of an EAS and the chosen functional dependence represents an empirical description of the data.
The effect results from two contributions with increasing zenith angle: (a) the increasing  $\mu/$EM ratio, and (b) the increasing average muon track length in the detector (see Section 4).
Both effects imply that the same recorded signal in VEM units has been produced by a smaller number of particles and therefore is subject to larger ﬂuctuations.

\subsection{Systematic error on the reconstruction due to zenith angle dependence}

In order to have an indication of the inﬂuence of the present knowledge of the signal accuracy on the reconstruction accuracy of individual events, we studied the systematic effect on $S(1000)$ due to the zenith angle dependence of the signal accuracy using a sample of about 600 well-measured events.
Each event was reconstructed twice once using the average zenith angle signal accuracy (Eq.~(3)) and once using the explicit zenith angle dependence (Eq.~(4)) when constructing the $\chi^2$ to be minimized.

A comparison on the $S(1000)$ obtained using the average signal accuracy (Eq.~(3)) expression in the $\chi^2$ minimization, relative to the $S(1000)$ obtained using the explicit zenith angle dependence (Eq.~(4)) in the $\chi^2$ minimization shows that the $S(1000)$ value is underestimated in average by only $-2$\% for vertical showers and overestimated  by  about $+2$\% for quasi-horizontal shower.
The tails of the distribution extend up to about $\pm7$\%.
These tails constitute less than 1\% of the events.

\subsection{Trigger influence on the signal accuracy}

By requiring to have \emph{both} detectors triggered the signal accuracy gets distorted, as downward ﬂuctuations (below threshold) will not be recorded.
This triggering effect produces the damping of the signal accuracy below $\sim$10\,VEM.
A simple numerical method can be used to explore the signal accuracy behavior in the low signal range.

The method relies on the ﬂuctuations observed at large signals -- where there is no trigger efficiency influence -- and extrapolates them to lower signals.
The ﬁrst step is to simulate the incident signal distribution, which is done using a power-law distribution, in agreement with the measured signal spectra in the stations.
After that, the signal distribution is sampled and the signal is ﬂuctuated twice following Eq.~(3).
Using the method described in Section 2 the signal accuracy can be obtained for the simulated data.
By introducing a threshold condition the ﬂuctuations damping is reproduced and compared with data.
The simulated and experimental ﬂuctuations agree within about 15\%, thus proving that the behavior of the signal accuracy obtained for large signals can be extrapolated to the low signal range, and that the interpretation of the data in such region as due to trigger effects is indeed correct.

\begin{figure}[t]
\centering
\includegraphics[width=\columnwidth]{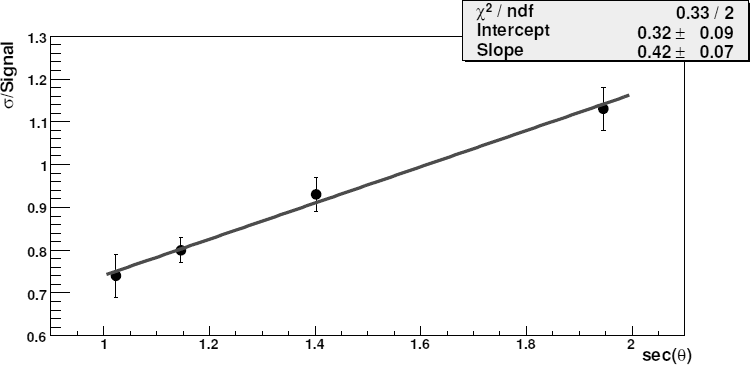}
\caption{Signal accuracy zenith angle dependence.
See text for details.}
\end{figure}

\section{Discussion}

To have a better understanding of the experimental data, the detector response has been fully simulated using the Auger Ofﬂine software featuring GEANT4 for the simulation of the SD [10].
Most of the particles that reach ground level are muons, electrons and gammas (both being  detected through the EM cascades they induce in water,  with the electrons producing Cherenkov light besides the small EM cascade).
Typical energies are a few GeV and a few MeV, for muons and EM particles, respectively.
We simulated 3000 individual particles impinging the detector at random positions on the tank for four ﬁxed zenith angles (0$^{\circ}$, 15$^{\circ}$, 30$^{\circ}$, and 60$^{\circ}$).

The signal measured by the PMTs is proportional to the amount of Cherenkov light produced, which in turn is proportional to the particle track length inside the detector.
Given the large energy difference between the muons and the EM component, and their interactions inside the detector, the response of the Auger tanks is different for  the different species.

The water is approximately 120\,cm thick (about three radiation lengths at the typical electron and gamma ground level energies) and thus the electromagnetic component is measured calorimetrically.
As the EM component is absorbed within about 10\,cm, almost any trajectory will  assure the particle and its cascade will stop inside the detector, being therefore basically independent of the  zenith angle and with small ﬂuctuations for ﬁxed geometry (of the order of the Poissonian ones).

Concerning muons, the response of the detector is proportional to their total track length in the tank [11].
Our simulation shows that a muon hitting a tank on a random position (either the top or the side) increases its signal by about 40\%, from vertical to 60.
Since our signal is measured in units of VEM,\footnote{A VEM corresponds to about 100\,p.e./PMT, and the p.e.\ yield for EM particles in the MeV energy range is of the order of 1\,p.e./MeV.} the same signal at large zenith angles will be due to a smaller number of particles, with larger inﬂuence of Poissonian ﬂuctuations.

Moreover, the ﬂuctuations on inclined muons are dominated by the changes in the total track length of the particle inside the tank (corner clipping muons).
This effect is driving the increase in fluctuations from about $3\times\sqrt{N_\text{pe}}$ for vertical to about $10\times\sqrt{N_\text{pe}}$ for 60$^{\circ}$.

The standard deviation in Eq.~(4) appears to be less than $\sqrt S$, that is below Poisson, for showers near the vertical.
This is an artifact of the units we use.
The measurement is made in VEM, as explained above.
For showers near vertical there is a high proportion of EM particles in the showers.
On account of their relatively low energies many EM particles are required to deposit one VEM in the detector, and EM particles generally deposit all their energies in the tank. Hence in these units their ﬂuctuations appear low.

\section{Conclusions}

The Auger Surface Detector signal accuracy has been measured using data from EAS and can be modeled by a Poissonian term times a normalization constant related to the primary cosmic ray zenith angle.

From the data, and with the validity limits of having non-saturated signals beyond 200\,m from the shower core, the following expression have been obtained (signals being expressed in VEM units):
\begin{equation}
   \sigma  = [(0.99 \pm 0.02) \times \sqrt{s}],
\end{equation}
which folds in the zenith angle distribution of events.
A explicit dependence on the shower axis’ zenith angle was measured up to 68$^\circ$:
\begin{equation}
   \sigma  = [(0.32 \pm 0.09) + (0.42 \pm 0.07) \times \sec(\theta)] \times \sqrt{s}.
\end{equation}


\section*{References} 

\begin{sloppypar}
\footnotesize\setlength{\parindent}{0mm}\setlength{\parskip}{2mm}

\noindent
[1]	The Pierre Auger Collaboration, Layout of the Pierre Auger Observatory, Proceedings of 27th ICRC, Hamburg, Germany, Copernicus Gesellschaft, HE, 2001, pp.~703--706.

\noindent
[2]	A.~Hillas, Proceedings of 11th ICRC, Budapest, Hungary, Acta Phys.\ Acad.\ Sci.\ Hung.\ 29 (Suppl.\ 3) (1970) 355.

\noindent
[3]	The Pierre Auger Collaboration, Nucl.\ Instr.\ Meth.\ A 523 (2004) 50--95.

\noindent
[4]	The Pierre Auger Collaboration, Surface Detector Construction and Installation at the Auger Observatory, The AUGER Collaboration, Proceedings of 27th ICRC, Hamburg, Germany, Copernicus Gesellschaft, HE, 2001, pp.~749--752.

\noindent
[5]	The Pierre Auger Collaboration, Calibration of the surface array of the Pierre Auger Observatory, M.~Aglietta, et al., for the Pierre Auger Collaboration, Proceedings of 29th ICRC, Pune, India (usa-allison-ps-abs1-he14).

\noindent
[6]	Pierre Auger Collaboration contributions to 27th ICRC, (Copernicus Gesellschaft, Hamburg, Germany 2001) and 28th ICRC (Universal Academy Press, Tokyo, Japan, 2003).

\noindent
[7]	Particle Data Group, Phys.\ Lett.\ B 592 (2004) 279.

\noindent
[8]	The Pierre Auger Collaboration, The trigger system of the PAO surface detector: operation, efﬁciency and stability, D.~Allard, et al., for the Pierre Auger Collaboration, Proceedings of 29th ICRC, Pune,India (usa-lhenry-yvon-I-abs1-he14-poster).

\noindent
[9]	The Pierre Auger Collaboration, Signal ﬂuctuations in the Auger surface   detector,   Proceedings   of   28th   ICRC,    HE-1,    2003, pp.\ 469--472.

\noindent
[10] The Pierre Auger Collaboration, The ofﬂine software framework of the Pierre Auger Observatory, Proceedings of 29th ICRC, Pune,  India (usa-paul-T-abs1-he15-poster).

\noindent
[11] A.~Etchegoyen, et al., Nucl.\ Instr.\ Meth.\ A 545 (2005) 602.
\end{sloppypar}

\end{document}